\documentstyle[prb,aps,multicol,epsf]{revtex}

\begin{document}

\draft
\title{
Chiral ordered phases in a frustrated $S=1$ chain 
with uniaxial single-ion-type anisotropy
}
\author{Toshiya Hikihara$^*$}
\address{Department of Earth and Space Science, Faculty of Science, 
Osaka University, Toyonaka 560-0043, Japan}
\date{June 11, 2000}
\maketitle
\begin{abstract}
The ground-state phase transitions of a frustrated $S = 1$ Heisenberg chain 
with the uniaxial single-ion-type anisotropy and 
the frustrating next-nearest-neighbor coupling are studied.
For the system, it has been shown that there are 
gapless and gapped chiral phases 
in which the chirality $\kappa_l = S^x_l S^y_{l+1} - S^y_l S^x_{l+1}$ 
exhibits a finite long-range order (LRO) 
and the spin correlation decays either algebraically or exponentially.
In this study, the transitions between the Haldane and chiral phase
and between the large-D (LD) and chiral phase are investigated 
using the infinite-system density-matrix renormalization group method.
It is found that there exist two types of gapped chiral phases, 
^^ ^^ chiral Haldane" and ^^ ^^ chiral LD" phases, 
in which the string LRO coexists with the chiral LRO and 
the string correlation decays exponentially, respectively.
\end{abstract}

\begin{multicols}{2}

Frustrated quantum spin chains and their low-energy properties 
have been one of the most attractive subjects over the decades.
In the systems, a rich variety of magnetic phases is realized 
by the interplay between frustration and quantum fluctuation.
More recently, both analytic~\cite{Ner,Kole} and 
numerical~\cite{Kabu,Hiki} studies have been performed 
to investigate the possibility of a new phase 
defined by the nonzero value of the $z$-component 
of the total vector chirality~\cite{vecchl} 
and the absence of the helical long-range order (LRO)
\begin{eqnarray}
\kappa &=& \frac{1}{L} \sum_l \kappa_l \ne 0, \label{eq:chl} \\
& &\kappa_l = S^x_l S^y_{l+1} - S^y_l S^x_{l+1}
          = \left[ \vec{S}_l \times \vec{S}_{l+1} \right]_z, 
\label{eq:vecchl} \\
\vec{m}(q) &=& \frac{1}{L} \sum_l \vec{S}_l {\rm e}^{iql} = 0~~~~~ 
({\rm for} ~\forall ~q),
\end{eqnarray}
where $\vec{S}_l$ is spin-S operator at the $l$-th site and 
$L$ is the total number of spins.
This chiral ordered phase is characterized by 
the spontaneous breaking of the parity symmetry 
with preserving the time-reversal symmetry.
From the studies, it has been revealed that the frustrated $XXZ$ 
spin chains with the competing nearest- and next-nearest-neighbor couplings 
exhibit the chiral LRO for the cases of 
$S = 1/2$, $1$, $3/2$, and $2$.~\cite{arbS}
The ground-state phase diagrams and the critical properties of the chains 
has also been investigated.

One of the most surprising observations among the studies 
is that there exist two different types of chiral phases 
in the frustrated $S = 1$ $XXZ$ chain.~\cite{Kabu,Hiki}
It has been shown numerically that, 
in the one of them, the ^^ ^^ chiral Haldane" phase, 
the chiral and string LRO's coexist and 
the spin correlation decays exponentially with a finite energy gap, 
whereas in the other, the ^^ ^^ gapless chiral" phase, 
the chiral LRO exists and the string and spin correlations decay 
algebraically with gapless excitations.
The appearance of the chiral phase with gapfull excitations 
has been also predicted for arbitrary integer $S$ 
by the field theoretical approach.~\cite{Kole}

The aim of this paper is to study the ground-state phase transitions of 
the frustrated $S = 1$ Heisenberg chain 
with an single-ion-type anisotropy,
\begin{equation}
{\cal H} = \sum_{\rho=1,2} J_\rho \sum_l \vec{S}_l \cdot \vec{S}_{l+\rho} 
         + D \sum_l \left( S_l^z \right)^2,    \label{eq:Ham}
\end{equation}
in which the appearance of the chiral LRO has been predicted~\cite{Kole} 
as well as the frustrated $XXZ$ chains.
Hereafter I consider the case of 
$j \equiv J_2/J_1 > 0$ and $d \equiv D/J_1 > 0$.
In the system of the case, it has been shown that 
there exist three phases besides the chiral phases; 
the Haldane phase characterized by the string LRO, 
the large-D (LD) phase, 
and the double-Haldane phase in which the system 
can be regarded as two Haldane subchains coupled weakly.~\cite{DH1,DH2}
It is therefore natural to expect that there are phase transitions 
between the chiral phases and the other phases.
In this work, I study the ground-state phase transitions 
between the Haldane and chiral phases 
and between the LD and chiral phases.
In both regions, 
I have found that the system undergoes two successive transitions, 
first from the Haldane or LD phase to a chiral phase with a finite gap,
and then, from the gapped chiral phase to the gapless chiral phase.
It has been also found that the gapped chiral phases are 
distinct from each other; the one is the chiral Haldane phase 
with a finite string LRO, while the other is the ^^ ^^ chiral LD" phase 
with no string LRO.

In order to investigate the nature of the transitions in detail, 
I compute the two-point chiral, string, and spin correlation functions 
defined by
\begin{eqnarray}
C_\kappa (r) &=& \langle \kappa_{l_0-r/2} \kappa_{l_0+r/2} \rangle, 
\label{eq:Cchl} \\
C_{\rm str} (r) &=& \langle S^z_{l_0 - r/2} 
     \exp \left( i \pi \sum_{j=l_0-r/2}^{l_0+r/2-1} S^z_j \right)
        S^z_{l_0+r/2} \rangle,   \label{eq:Cstr} \\
C^x_s (r) &=& \langle S^x_{l_0-r/2} S^x_{l_0+r/2} \rangle, 
\label{eq:Cx}
\end{eqnarray}
which are associated with the chiral, string, and helical order parameters, 
respectively.
Here $l_0$ represents the center position of the open chain, i.e., 
$l_0 = L/2$ for even $r$ and $l_0 = (L+1)/2$ for odd $r$.
For the transitions between the Haldane and chiral phases 
(the LD and chiral phases), I calculate these correlations 
for $d=0.6$ ($j = 0.7$) and various values of $j$ ($d$).
And then, I determine the phase diagram on each line 
by examining the behavior of the correlations at long distance.
The calculation has been performed using the density-matrix 
renormalization group (DMRG) method proposed by White.~\cite{White1,White2}
I employed the infinite-system algorithm of the accelerated version 
proposed by Nishino and Okunishi.~\cite{PWF1,PWF2}
The number of kept states $m$ is up to $350$ and the $m$-convergence of 
the data has been checked by increasing $m$ consecutively.

\begin{figure}
\narrowtext
\begin{center}
\noindent
\leavevmode\epsfxsize=72mm
\epsfbox{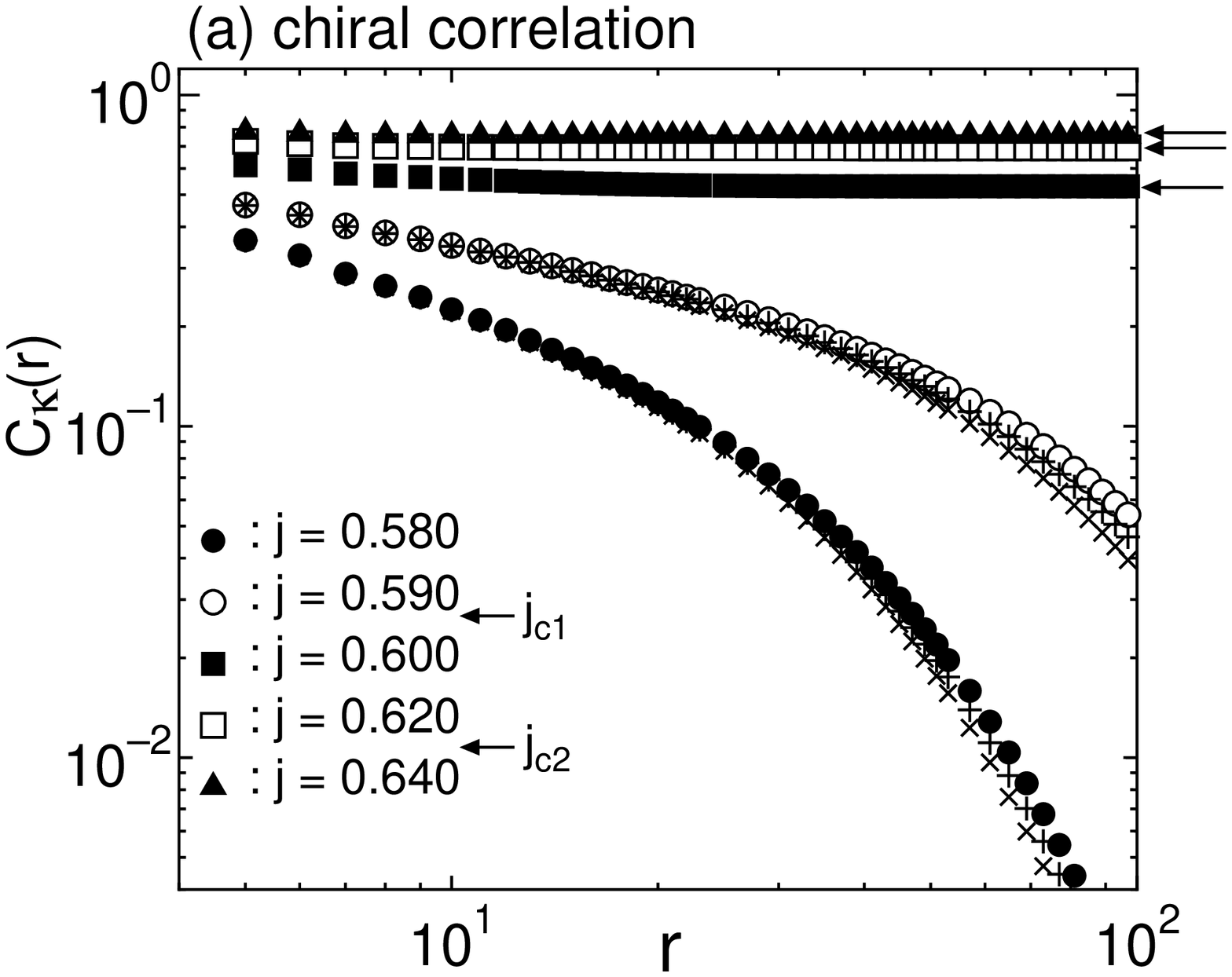}
\end{center}
\vspace{-0.3cm}
\begin{center}
\noindent
\leavevmode\epsfxsize=72mm
\epsfbox{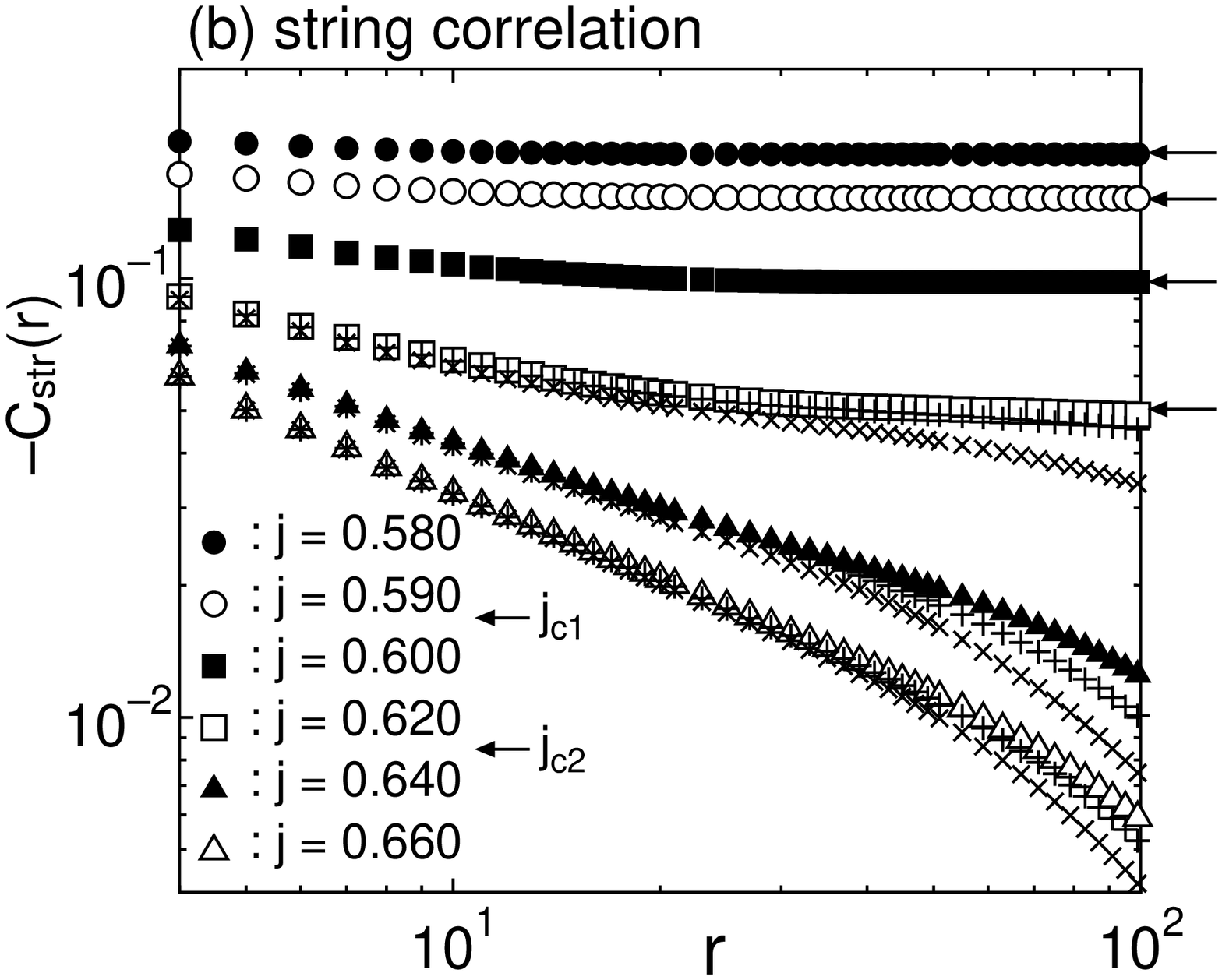}
\end{center}
\vspace{-0.3cm}
\begin{center}
\noindent
\leavevmode\epsfxsize=70mm
\epsfbox{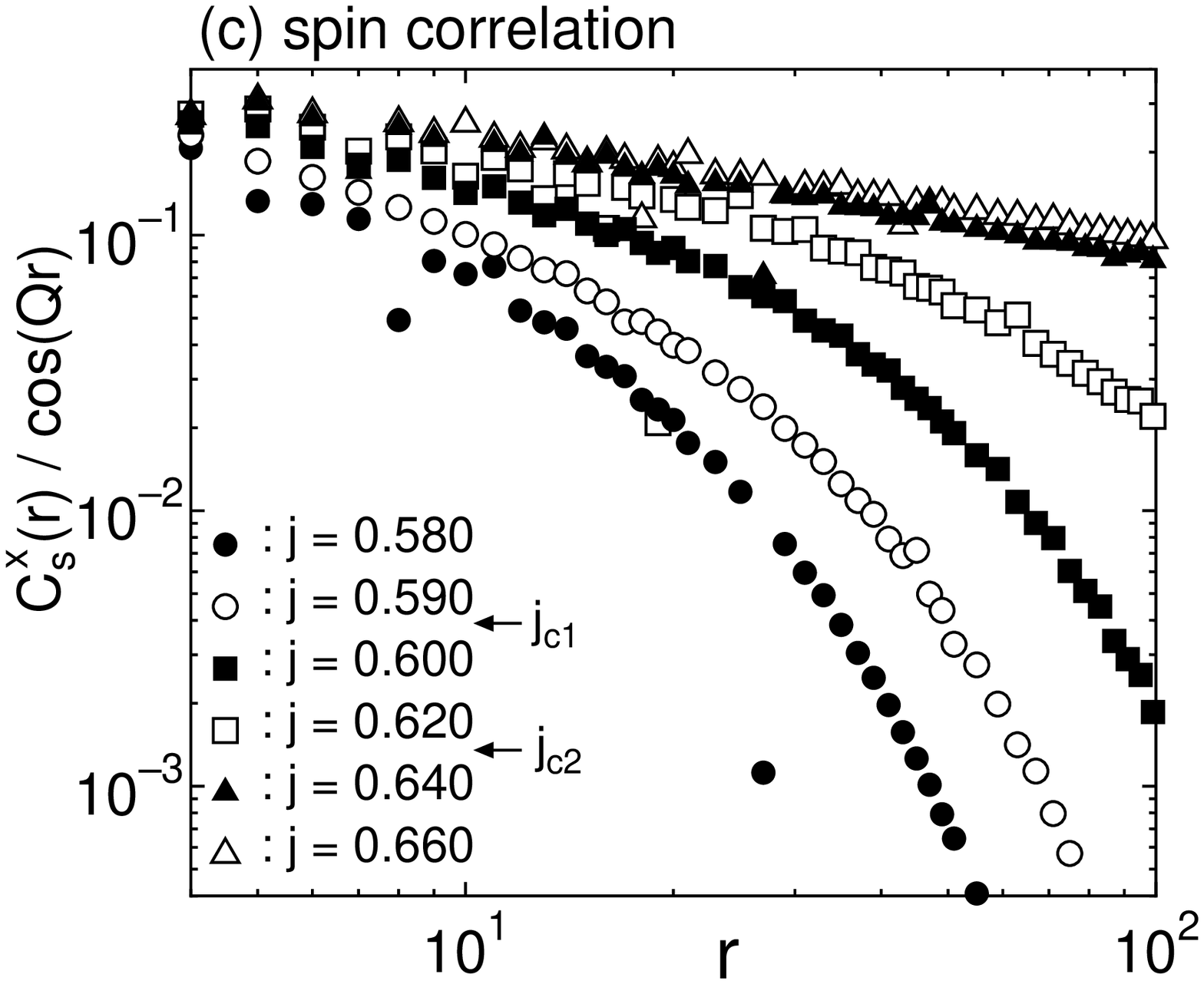}
\end{center}
\caption{The $r$-dependence of the correlation functions 
   for $d = 0.6$: 
   (a) chiral correlation $C_\kappa(r)$; 
   (b) string correlation $-C_{\rm str}(r)$; 
   (c) spin correlation $C_{\rm s}^x(r)$ divided 
   by the oscillating factor ${\rm cos}(Qr)$.
   (In figure (a), the data of $C_\kappa(r)$ for $j = 0.660$, 
   which are almost the same as those for $j = 0.640$, 
   are omitted for clarity.)
   The number of kept states is 
   $m = 260$ (for $j = 0.580$), 
   $m = 300$ (for $0.590 \le j \le 0.640$), 
   and $m = 350$ (for $j = 0.660$).
   To illustrate the $m$-dependence, I also indicate the data 
   with smaller $m$ by crosses for several cases. 
   In the other cases, the numerical errors of the data are smaller 
   than the symbols.
   Arrows in the figures represent the extrapolated $r=\infty$ values.}
\label{fig:H-C}
\end{figure}

Let us see first the results for the Haldane-chiral transition.
Figure \ref{fig:H-C} shows the $r$-dependence of the calculated 
chiral, string, and spin correlation functions 
for $d = 0.6$ and several typical values of $j$ 
on the $\log$-$\log$ plots.
It can be clearly seen in Fig. \ref{fig:H-C} (a) that 
the chiral correlation $C_\kappa (r)$ is bent upward for larger $r$ 
for $j > j_{c1} = 0.595 \pm 0.005$ indicating a finite chiral LRO, 
whereas it is bent downward for $j < j_{c1}$ 
indicating an exponential decay.
In Fig. \ref{fig:H-C} (b), on the other hand, 
the data of $C_{\rm str} (r)$ for $j < j_{c2} = 0.630 \pm 0.010$
curve upward for larger $r$ indicating a finite string LRO, 
while they exhibit a linear behavior for $j > j_{c2}$ 
indicating a power-law decay.~\cite{trunc}
It should be noticed that the estimate of $j_{c2}$ is 
distinctly larger than that of $j_{c1}$, 
and accordingly, there is an intermediate phase where 
both the chiral and string LRO's coexist 
between the Haldane phase and the chiral phase without the string LRO.
This observation that there are different types of chiral phases 
is also confirmed by the behavior of the spin correlation 
$C_s^x (r)$ in Fig. \ref{fig:H-C} (c); 
It decays exponentially for $j < j_{c2}$ with a finite gap 
whereas it decays algebraically with gapless excitations.
Thus, it can be concluded that, as $j$ increases, 
the system undergoes two successive transitions, 
first at $j = j_{c1}$ from the Haldane phase to 
the ^^ ^^ chiral Haldane" phase characterized by 
the coexistence of the chiral and string LRO's and a finite gap, 
and then, at $j = j_{c2}$ from the chiral Haldane phase 
to the gapless chiral phase.
Here I note that this results on the appearance of 
the two successive transitions and the nature of the intermediate phase 
are the same as those found on 
the frustrated $S=1$ $XXZ$ chain.~\cite{Kabu,Hiki}

\begin{figure}
\narrowtext
\begin{center}
\noindent
\leavevmode\epsfxsize=72mm
\epsfbox{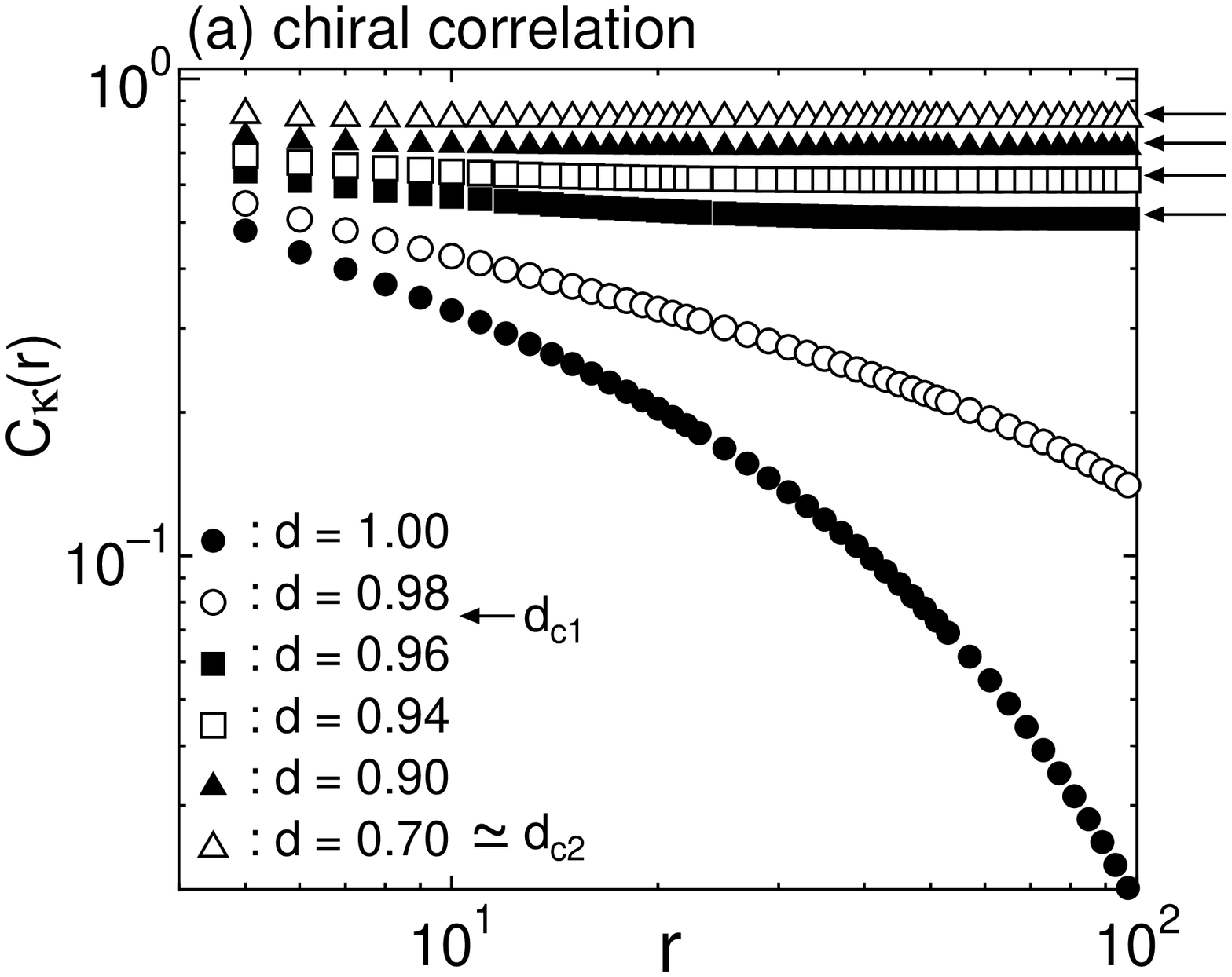}
\end{center}
\vspace{-0.3cm}
\begin{center}
\noindent
\leavevmode\epsfxsize=70mm
\epsfbox{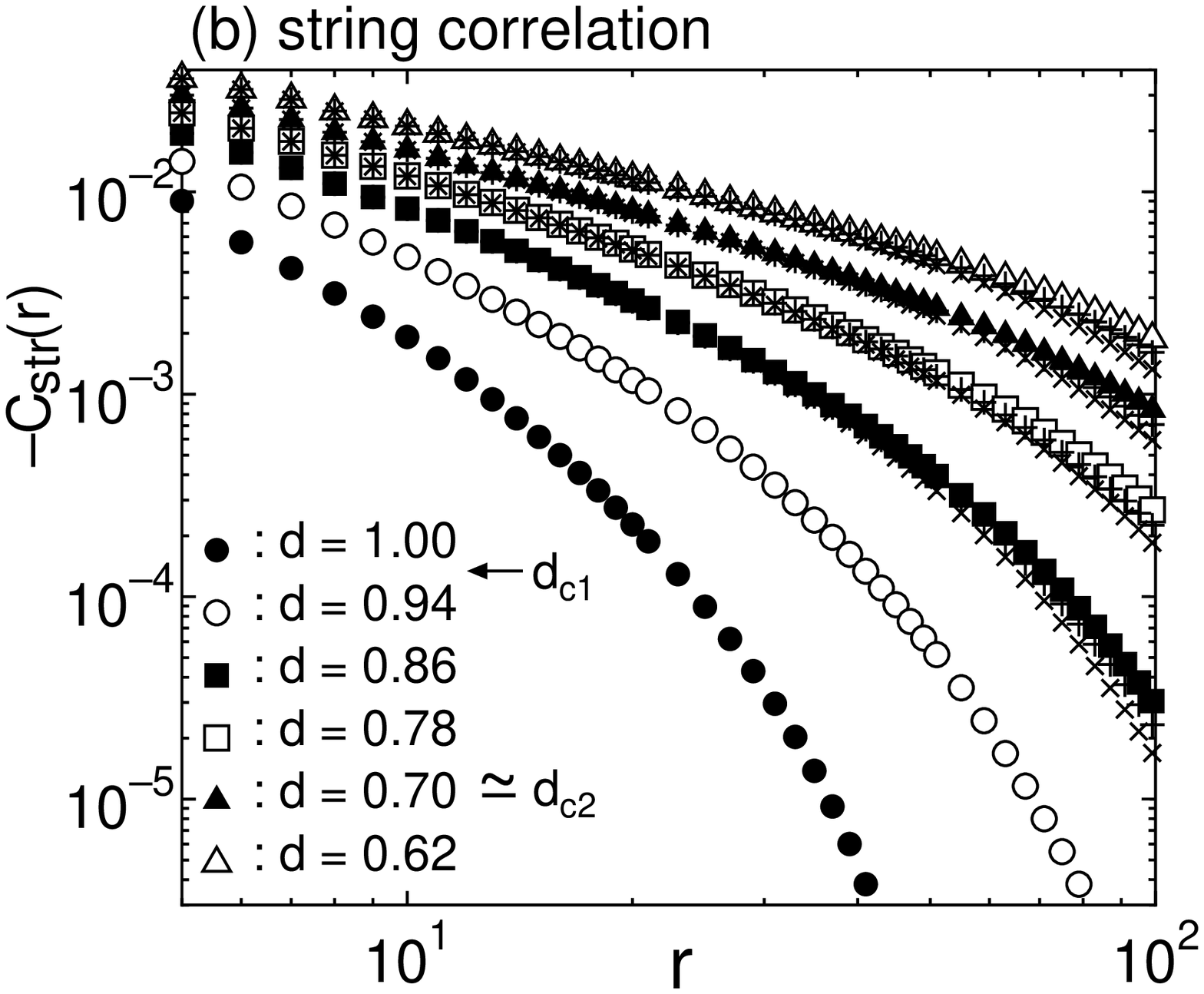}
\end{center}
\vspace{-0.3cm}
\begin{center}
\noindent
\leavevmode\epsfxsize=70mm
\epsfbox{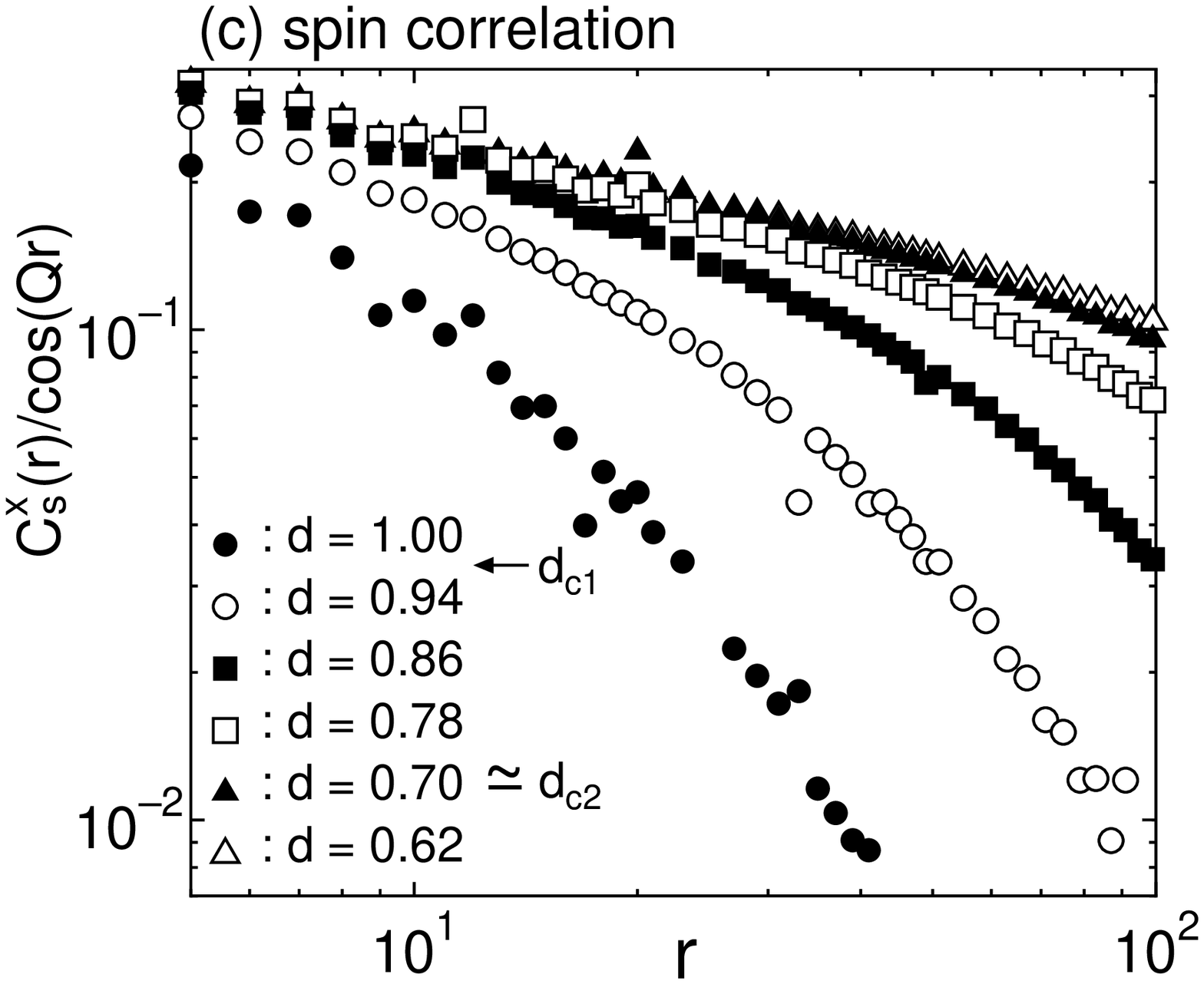}
\end{center}
\caption{The $r$-dependence of the correlation functions 
   for $j = 0.7$: 
   (a) chiral correlation $C_\kappa(r)$; 
   (b) string correlation $-C_{\rm str}(r)$; 
   (c) spin correlation $C_{\rm s}^x(r)/{\rm cos}(Qr)$.
   (Note that the selected values of $d$ for figure (a) are 
   different from those for figure (b) and (c).)
   The number of kept states is $m = 300$ (for $d \ge 0.86$) 
   and $m = 350$ (for $d \le 0.78$).
   For several cases, the data with smaller $m$ are also indicated 
   by crosses.
   The numerical errors in the other cases are smaller 
   than the symbols.
   Arrows in the figures represent the extrapolated $r=\infty$ values.}
\label{fig:LD-C}
\end{figure}

Next, we consider the phase transitions between 
the LD and chiral phases.
In Fig. \ref{fig:LD-C}, I show the correlation functions 
for $j = 0.7$ and several typical values of $d$ on the $\log$-$\log$ plots.
As shown in the figure, the chiral correlation $C_\kappa (r)$ 
exhibits a finite LRO for $d < d_{c1} = 0.97 \pm 0.01$, 
while it shows an exponential decay for $d > d_{c1}$.
In the meantime, the spin and string correlations 
decay exponentially for $d > d_{c2} = 0.70 \pm 0.08$, 
whereas they decay algebraically for $d < d_{c2}$.
Here it should be pointed out that the estimate of $d_{c2}$ 
has a quite large error bar due to 
the truncation error of the DMRG calculation, 
which tends to underestimate the correlation length.
From the behavior of the correlations for $0.86 \le d < d_{c1}$ in the figure, 
however, it can be clearly seen that 
the gapped 

\noindent
region with a finite chiral LRO does exist, and consequently, 
the value of $d_{c2}$ is distinctly smaller than that of $d_{c1}$.
It is therefore concluded that, 
as well as the Haldane-chiral transitions, 
the system undergoes two successive transitions as $d$ decreases, 
first from the LD to the gapped chiral phase at $d = d_{c1}$ 
and then from the gapped to the gapless chiral phase at $d = d_{c2}$.
This gapped chiral phase 
between the LD and gapless chiral phases, the ^^ ^^ chiral LD" phase, 
is different from the chiral Haldane phase 
in the respect that it has no string LRO, 
and accordingly, it can be regarded as a new type 
of the gapped chiral phases.

In summary, I have investigated numerically 
the ground-state phase transitions between the Haldane and chiral phases 
and between the LD and chiral phases 
in the frustrated $S=1$ Heisenberg spin chain 
with a single-ion-type anisotropy.
For the Haldane-chiral transition, I have found that 
the system undergoes two successive transitions 
which behavior is similar to the ones found in 
the frustrated $S = 1$ $XXZ$ chain~\cite{Kabu,Hiki}: 
First, the system moves from the Haldane phase to the chiral Haldane phase, 
and then, moves to the gapless chiral phase.
For the LD-chiral transition, it has been found again that, 
as $d$ decreases, there occur two successive transitions, 
the one from the LD phase to a chiral phase with a finite gap 
and the other from the gapped chiral to the gapless chiral phase.
The gapped chiral phases between the LD and gapless chiral phases 
has no string order and is distinct from the chiral Haldane phase.
The full phase diagram on the $j$-$d$ plane will be reported elsewhere.

The author thanks M. Kaburagi and H. Kawamura 
for fruitful discussions.
This work was supported by the Japan Society 
for the Promotion of Science for Young Scientists.

\end{multicols}

\begin{thebibliography}{99}
\bibitem{vecchl}
Note that the vector chirality (\ref{eq:vecchl}) differs from 
the scalar chirality defined by 
$\chi_l = \vec{S}_{l-1} \cdot \vec{S}_l \times \vec{S}_{l+1}$, 
which is often discussed in studies on isotropic spin systems.
\bibitem{trunc}
The data of the string correlation for $j > j_{c2}$ 
slightly deviate from the linear behavior for larger $r$.
However, from the $m$-dependence of the data, I believe that 
it is the truncation error of the DMRG calculation and 
not a serious problem.
\end{thebibliography}
\end{document}